\documentstyle[12pt]{article}
\renewcommand{\epsilon}{\varepsilon}
\begin{document}
\title{Conformal symmetry and  deflationary gas universe}
\author{Winfried Zimdahl\\
Universit\"at Konstanz, 
D-78457 Konstanz, Germany\\
and\\
Alexander B. Balakin \\ 
Kazan State University, 420008 Kazan, Russia}
\date{December 1, 2000}
\maketitle
\begin{abstract}
We describe the ``deflationary'' evolution from 
an initial de Sitter phase to a subsequent Friedmann-Lema\^{\i}tre-Robertson-Walker 
(FLRW) period as a specific non-equilibrium configuration of a 
self-interacting gas. 
The transition dynamics corresponds to a conformal, timelike symmetry of an ``optical'' metric, characterized by a refraction index of the cosmic 
medium which continously decreases from a very large initial value to unity in the FLRW phase. 
\end{abstract}  

\section{Introduction}

Thermodynamic equilibrium in the expanding universe is  only possible under special circumstances. 
For a simple gas model of the cosmic medium e.g., the  
relevant equilibrium condition can be satisfied only for massless particles, corresponding to an equation of state $P=\rho /3$,  
where $P$ is the pressure and $\rho $ is the energy density \cite{Stew,Ehl}. 
This equilibrium condition is equivalent to a symmetry requirement for the cosmological dynamics: 
The quantity $u ^{a}/T$, where $u ^{a}$ is the four velocity of the medium and $T$ is its temperature, has to be a conformal Killing vector (CKV) 
\cite{TauWei,Cher}. 
For deviations from $P=\rho /3$ the conformal symmetry is violated and no equilibrium is possible. 
If we give up the diluted gas approximation on which this model is based and take into account additional interactions inside the many-particle system, the situation changes. 
Suitable interaction terms give rise to a generalization of the equilibrium conditions of the gas (see  
\cite{ZiBa3} and references therein). 
Among these generalized equilibrium configurations are ``true'' equilibrium states, i.e. states with vanishing entropy production, but there are also specific types of non-equilibrium states. 
The latter are characterized by an equilibrium distribution function, the moments of which are not conserved, however.  
The corresponding source terms in the balance equations for the moments are connected with the production of particles and entropy. 
They represent not necessarily small deviations from ``standard'' equilibrium which are of interest for the investigation of cosmological out-of-equilibrium processes. 

Gas models of the cosmic medium have the advantage that the relation between microscopic dynamics (kinetic theory) and a macroscopic thermo-hydrodynamical 
description is exactly known. 
An apparent disadvantage of these models is that they are restricted to 
``ordinary'' matter with equations of state in the range 
$0 < P \leq \rho /3$.  
However, periods of accelerated  expansion, either inflation or a dark energy dominated dynamics, require  ``exotic'' matter, namely a substratum with an effective negative pressure that violates the strong energy condition $\rho +3P > 0$.   
Usually, such kind of matter is modeled by the dynamics of scalar fields with suitable potentials. 
Here we show that the generalized equilibrium concept provides a framework 
which makes use of the advantages of gas models and at the same time avoids the restriction to equations of state for ``ordinary'' matter. 
The point is that the specific non-equilibrium configurations mentioned above may give rise to a negative pressure of the cosmic medium. 
If the latter is sufficiently large to violate the strong energy condition, it induces an accelerated expansion of the universe. 
Our main interest here is to understand the 
``deflationary'' \cite{Barrow} transition from an initial de Sitter phase to a subsequent 
FLRW period as a specific non-equilibrium configuration which is based on the 
generalized equilibrium concept. 
Moreover, we show that the non-equilibrium contributions may be mapped onto an refraction index $n _{r}$ of the cosmic medium which is used to define an ``optical''  
metric $\bar{g} _{ik}=g _{ik}+\left[1-n _{r}^{-2} \right]u _{i}u _{k}$ 
(cf. \cite{Gordon,Ehlopt}). 
The circumstance that this kind of non-equilibrium retains essential equilibrium aspects is reflected by the fact that 
$u ^{a}/T$ is a CKV of the optical metric 
$\bar{g}_{ik}$ during the entire transition period.  
As the FLRW stage is approached, the refraction index tends to unity  
which implies $\bar{g}_{ik} \to g _{ik}$.

\section{Fluid description}

Let the fluid be characterized by a four velocity $u ^{i}$ and a temperature $T$. 
Under the conditions of spatial homogeneity and isotropy the Lie derivative $\pounds _{\xi }g _{ik}$ of the metric 
$g _{ik}$ with respect to the vector $\xi ^{a}\equiv  \frac{u ^{a}}{T}$ may be decomposed  
into contributions parallel and perpendicular to the four-velocity: 
\begin{equation}
\pounds _{\xi } g _{ik} \equiv  
\left(\frac{u _{i}}{T}\right)_{;k} 
+ \left(\frac{u _{k}}{T}\right)_{;i}
= {\rm 2} A u _{i}u _{k} + {\rm 2} \phi h _{ik}  \ ,
\label{1}
\end{equation}
where 
$h ^{ik}=g ^{ik}+u ^{i}u ^{k}$.  Units are chosen so that $c=k _{B}=1$.
With the help of the projections 
\begin{equation}
\frac{\dot{T}}{T} = AT = 
\frac{T}{2}u ^{i}u ^{k}\pounds _{\xi }  g _{ik}\ ,\quad \quad 
{\rm \Theta } = {\rm 3}T \phi = \frac{T}{{\rm 2}}h ^{ik}
\pounds _{\xi } g _{ik} \ ,
\label{2}
\end{equation}
one obtains the rate of change of the temperature $\frac{\dot{T}}{T}$, 
where $\dot{T} \equiv  T _{,a}u ^{a}$, 
and the fluid expansion $\Theta \equiv  u ^{a}_{;a}$, respectively. 
$\frac{\dot{T}}{T}$ and $\Theta $ 
are given by orthogonal projections of the Lie derivative. 
However, in a cosmological context one expects a relationship between these two quantities which represents the cooling rate of the cosmic fluid in an expanding universe. 
To establish this link we take into account that the fluid is characterized by a particle flow vector $N ^{i}$ and an energy-momentum tensor 
$T^{ik}_{\left(eff \right)}$:  
\begin{equation}
N ^{i} = n u ^{i}\ ,
\quad  
T^{ik}_{\left(eff \right)} = \rho u^{i}u^{k} 
+ P h^{ik}\ ,\quad P = p + \Pi  \ .
\label{3}
\end{equation}
Here, $n$ is the particle number density, $\rho $ is the energy density seen by a comoving observer, $P$ is the total pressure with the equilibrium part $p$ and 
the non-equilibrium contribution $\Pi $. 
The balance equations are
\begin{equation}
N^{a}_{;a}  =  
\dot{n} + 3H n =  n \Gamma  \ , \quad \quad 
- u _{i}T ^{ik}_{\left(eff \right);k}=\dot{\rho } + 3H\left(\rho + P \right)
= 0 \ , 
\label{4}
\end{equation}
where $\Gamma $ is the particle production rate, i.e., we have admitted the possibility of particle number non-conserving processes. Moreover, we have introduced the Hubble rate $H$ by $\Theta \equiv  3H$.  
Furthermore, we assume equations of state in the general form 
$p = p \left(n,T \right)$ and 
$\rho  = \rho  \left(n,T \right)$, 
i.e., particle number density and temperature are the independent thermodynamical variables. 
Differentiating the latter relation, using the balances (\ref{4})  and restricting ourselves to constant entropy per particle, we obtain the cooling rate 
\begin{equation}
\frac{\dot{T}}{T} = - 3H \Delta  
\left(\frac{\partial p}{\partial \rho } \right)_{n} \ ,
\quad \quad 
\Delta \equiv  1 - \frac{\Gamma }{3H } \ . 
\label{6}
\end{equation}
Comparing the expressions in (\ref{2}) and (\ref{6}) for the temperature evolution, we find that they are consistent for 
\begin{equation}
\pounds _{\xi } g _{ik}  =
\frac{{\rm 2H}}{T}\left\{g _{ik} 
+ \left[{\rm 1} - {\rm 3} {\rm \Delta } 
\left(\frac{\partial{p}}{\partial{\rho }} \right)_{n} \right] 
u _{i}u _{k}\right\}  \ .
\label{7}
\end{equation}
In the special case 
$\Delta =1$, $p=\frac{\rho }{3}$ we recover that 
$\frac{u ^{a}}{T}$ is a CKV of the metric $g _{ik}$. 
Equation (\ref{7}) represents a specific modification of the  CKV condition. 
Since the CKV property of $\frac{u ^{a}}{T}$ is known to be related to the dynamics of a standard radiation dominated FLRW universe, the question arises to what extent the replacement  
$g _{ik}  \rightarrow 
g _{ik} 
+ \left[1 - 3 \Delta \left(\partial{p}/\partial{\rho } 
\right)_{n} \right]
u _{i}u _{k}$ 
in (\ref{7}) 
gives rise to a modified cosmological dynamics. 
In particular, it is of interest to explore whether this modification admits phases of accelerated expansion of the universe.  
To clarify this question  we recall that the CKV property of 
$\frac{u ^{a}}{T}$  may be obtained as a ``global'' equilibrium condition in relativistic gas dynamics. 
In the following we ask whether the more general case, corresponding to the mentioned modification of the conformal symmetry, can be derived in a gas dynamical context as well. 

\section{Gas dynamics}

Relativistic gas dynamics is based on Boltzmann's equation for the one-particle distribution function $f=f \left(x,p \right)$, 
\begin{equation}
p ^{i}\frac{\partial{f}}{\partial{x ^{i}}} - \Gamma ^{k}_{il}p^{i} p^{l}
\frac{\partial{f}}{\partial{p^{k}}} 
+ mF^{i} \frac{\partial{f}}{\partial{p^{i}}}
=C [f]\ , 
\label{10}
\end{equation}
where $C \left[f \right]$ is Boltzmann's collision integral. 
$F ^{i}=F ^{i}\left(x,p\right)$ is a not yet specified effective one-particle force. Let us restrict ourselves to the class of forces which admit  solutions 
of Boltzmann's equation that are 
of the type of  J\"uttner's distribution function 
$
f^{0}\left(x, p\right) = 
\exp{\left[\alpha + \beta_{a}p^{a}\right] }$, 
where $\alpha = \alpha\left(x\right)$ and 
$\beta_{a}\left(x \right)$ is timelike. For $f \rightarrow f ^{0}$ the collision integral vanishes: $C \left[f ^{0} \right]=0$.   
Substituting $f ^{0}$ into Eq. (\ref{10}), we obtain the condition
$p^{a}\alpha_{,a} +
\beta_{\left(a;b\right)}p^{a}p^{b}   
=  - m \beta _{i}F ^{i}$.  
For the most general expression of the relevant force projection,  
$\beta _{i}F ^{i} =  \beta _{i}F ^{i}_{a}\left(x \right)p ^{a} 
+ \beta _{i}F^{i}_{ab}\left(x \right)p ^{a}p ^{b}$ (cf. \cite{ZiBa3}), 
this condition  decomposes into the generalized equilibrium conditions 
\begin{equation}
\alpha _{,a} = - m \beta _{i}F _{a}^{i} \ ,
\mbox{\ \ }\mbox{\ \ }\mbox{\ \ }
\beta _{\left(k;l \right)} 
\equiv  \frac{1}{2}\pounds _{\beta } g _{kl}
= - m \beta _{i}F ^{i}_{kl}\ .
\label{14}
\end{equation}
With the identification $\beta _{a} \rightarrow \frac{u _{a}}{T}$ 
we may read off that a force with 
\begin{equation}
- m \beta _{i}F ^{i}_{ab}  = \frac{H}{T}
\left\{g _{ab} +  
\left[1 - 3 \Delta \left(\frac{\partial{p}}{\partial{\rho }} \right)_{n} 
\right]u _{a}u _{b} \right\} 
\label{15}
\end{equation}
reproduces the previous consistency relation (\ref{7}).  
The latter is recovered as generalized equilibrium condition for particles in a specific force field. 

One may interpret this feature in a way which is familiar from gauge theories: 
Gauge field theories rely on the fact that local symmetry requirements (local gauge invariance) necessarily imply the existence of additional interaction fields (gauge fields). 
In the present context we impose 
the ``symmetry'' requirement (\ref{7}) (below we clarify in which sense the modification of the conformal symmetry is again a symmetry). 
Within the presented gas dynamical framework this ``symmetry'' can only be realized if one introduces additional interactions, here described by an effective force field $F ^{i}$. Consequently, in a sense, this force field may be regarded as the analogue of gauge fields. 

In its quadratic part the force has the structure 
$m \beta _{i}F ^{i} \propto \linebreak
\left\{m ^{2} 
+ \left[3 \Delta \left(\partial{p}/\partial{\rho } \right)_{n} 
- 1\right] E ^{2}\right\}H/T $,
where $E \equiv  -u _{a}p ^{a}$ is the particle energy. 
(For a discussion of the linear part see \cite{ZiBa3}.) 
This force, which governs the particle motion, depends on macroscopic and  microscopic quantities. 
The macroscopic quantities determine the motion of the individual microscopic particles which themselves are the constituents of the medium. 
This means, the system is a gas with internal self-interactions. 

In a next step we calculate the macroscopic transport equations for the self-interacting gas. 
The first and second moments of the distribution function are 
\begin{equation}
N^{i} = \int \mbox{d}Pp^{i}f\left(x,p\right) \mbox{ , } 
\quad \quad 
T^{ik} = \int \mbox{d}P p^{i}p^{k}f\left(x,p\right) \mbox{ .} 
\label{17}
\end{equation}
While $N ^{i}$ may be identified with the corresponding quantity in
 Eq. (\ref{3}), the second moment $T ^{ik}$  does {\it not} coincide with the energy-momentum tensor in Eq. (\ref{3}). 
With $f \rightarrow f ^{0} $ we obtain 
\begin{equation}
N ^{a}_{;a}= 3nH \left(1- \Delta  \right)\ ,\quad 
u _{a}T ^{ab}_{\ ;b}= 
- 3H \left(\rho + p \right)\left(1- \Delta  \right) \ .
\label{19}
\end{equation}
For $\Delta \neq 1$ there appear ``source'' terms in the balances. 
The crucial point now is that these source terms may consistently be mapped 
on the effective viscous pressure $\Pi $ of a conserved energy-momentum tensor of the type of $T ^{ik}_{\left(eff \right)}$ in Eq. (\ref{3}),  
according to (cf. \cite{ZiBa3})
\begin{equation}
\Delta  = 1 + \frac{\Pi }{\rho + p} 
\quad\Rightarrow\quad
\dot{\rho } + 3H \left(\rho + p + \Pi  \right) =0\ . 
\label{20}
\end{equation} 
In the following the cosmic medium is assumed to be describable by an energy-momentum tensor $T ^{ik}_{\left(eff \right)}$ where $\Pi $  is related to $\Delta $ by the latter correspondence. 
 
\section{Conformal symmetry}

For a homogeneous, isotropic, and spatially flat universe, the relevant dynamical equations are
\begin{equation}
8 \pi G\rho = 3 H ^{2}\ ,\quad 
\dot{H}= - 4\pi G\left(\rho + p + \Pi  \right)
\quad\Rightarrow\quad
\frac{\Gamma }{3H} = 1 + \frac{2}{3 \gamma }
\frac{\dot{H}}{H ^{2}} \ ,
\label{21}
\end{equation}
where $\gamma =1+p/ \rho $. 
A non-vanishing viscous pressure, in our case equivalent to the creation of particles, back reacts on the cosmological dynamics. 
To solve the equations (\ref{21}), assumptions about the ratio $\Gamma /H$ are necessary. Generally, one expects that particle production is a phenomenon which is characteristic for early stages of the cosmic evolution, while lateron 
a standard FLRW phase is approached. 
In case $H$ is a decaying function of the cosmic time, this is realized 
by the simple ansatz $\Gamma /H \propto H$ with the help of which 
Eqs. (\ref{21}) for 
ultrarelativistic matter 
($\gamma =4/3$) may be integrated: 
\begin{equation}
\frac{\Gamma }{H}\propto H \quad\Rightarrow\quad 
H = 2\frac{a _{e} ^{2}}
{a ^{2} + a _{e} ^{2}}H _{e} 
\quad\Rightarrow\quad
\Delta = \frac{a ^{2}}{a ^{2}+a _{e}^{2}}
\ .
\label{22}
\end{equation}
The Hubble rate $H$ changes continuously from $H = 2 H _{e}$, equivalent to 
$a \propto \exp{\left[Ht \right]}$ at $a \ll a _{e}$, to 
$H \propto a ^{-2}$, equivalent to $a \propto t ^{1/2}$, the standard radiation-dominated universe, for $a \gg a _{e}$. 
For $a<a _{e}$ we have $\ddot{a}>0$, while 
$\ddot{a}<0$ holds for $a>a _{e}$.  
The value $a _{e}$ denotes that transition from accelerated to decelerated expansion, corresponding to  
$\dot{H}_{e} = - H _{e}^{2}$. 
For $a \rightarrow 0$ we have $\Delta \rightarrow 0$, i.e., 
$\Gamma \rightarrow 3H$,  while for $a \gg a _{e}$ we find 
$\Delta \rightarrow 1$, i.e., $\Gamma \rightarrow 0$. 

Now we ask, how such kind of scenario may be realized within the previously discussed gas dynamics. 
To this purpose we introduce the redefinition $m F^{i}\to F ^{i}$ of the force since the ultrarelativistic limit corresponds to $m \to 0$. 
From the second of the generalized equilibrium conditions (\ref{14}) 
with (\ref{15}) and with $\Delta $ from (\ref{22}) one obtains the relevant force term: 
\begin{equation}
\pounds _{\xi } g _{ik}  = 
\frac{{\rm 2} H}{T}
\left[g _{ik} + \frac{a _{e}^{{\rm 2}}}{a ^{{\rm 2}}
+a _{e}^{{\rm 2}}} u _{i}u _{k}\right]\ , 
\quad\Rightarrow\quad
u _{i}F ^{i} =
- H E ^{{\rm 2}}\frac{a _{e}^{{\rm 2}}}{a ^{{\rm 2}}
+a _{e}^{{\rm 2}}} \ .
\label{23}
\end{equation}
An internal interaction, described by this component of an effective one-particle force $F ^{i}$ which is self-consistently exerted on the microscopic constituents of the cosmic medium, realizes the deflationary dynamics (\ref{22}). 

The explicit knowledge of the self-interacting force allows us to consider the microscopic particle dynamics which corresponds to the deflationary scenario. 
The solution of the equation of motion for the particles, 
$\mbox{D}p ^{i}/\mbox{d}\lambda =F ^{i}$, 
is $E 
\propto T \propto \left[a _{e}^{2}/\left(a ^{2}+ a _{e}^{2} \right) \right]^{1/2}$, i.e., the equilibrium distribution is indeed maintained which proves the consistency of our approach \cite{ZiBa3}. 
The particle energy changes from 
$E = {\rm const}$ for $a \ll a _{e}$ to $E \propto a ^{-1}$ for $a \gg a _{e}$. 

The outlined framework represents 
an exactly solvable model of a deflationary transition 
from an initial de Sitter phase to a subsequent radiation dominated FLRW period, both macroscopically and microscopically.  
This  transition  is accompanied by a characteristic change of the ``symmetry'' 
condition (\ref{14}) with (\ref{15}) from 
\begin{equation}
\pounds _{\xi } g _{ik} = \frac{{\rm 2}H}{T}
h _{ik} \quad {\rm for}\quad 
a \ll a _{e} \quad {\rm to} \quad 
\pounds _{\xi } g _{ik} = \frac{{\rm 2}H}{T}
g_{ik} \quad {\rm for}\quad 
a \gg a _{e} \ . 
\label{26}
\end{equation}
In order to clarify in which sense the modification of the conformal symmetry is a symmetry again,  
we introduce the ``optical metric'' (cf. \cite{Gordon,Ehlopt})
\begin{equation}
\bar{g}_{ik} = g _{ik} + \left[1 - n _{r}^{-2} \right]u _{i}u _{k}\ ,
\quad
n _{r}^{2} = 1 + a _{e}^{2}/a ^{2}= \Delta ^{-1}\ ,
\label{28}
\end{equation}
where $n _{r}$ plays the role of a refraction index of the medium which  
in the present case changes monotonically from $n _{r}\to \infty$ for 
$a \to 0$ to $n _{r} \to 1$ for $a \gg a _{e}$.  
For $a=a _{e}$ we have $n _{r}\left(a _{e} \right)=\sqrt{2} $.  
Optical metrics are known to be helpful in simplifying the equations of light propagation in isotropic refractive media. 
With respect to optical metrics light propagates as in vacuum.  
Here we demonstrate that such type of metrics, in particular their symmetry properties, are also useful in relativistic gas dynamics. 
Namely, it is easy to realize that the first relation (\ref{23}) is equivalent to 
\begin{equation}
\pounds _{\xi }\bar{g}_{ik} 
= \frac{{\rm 2}H}{T}\bar{g}_{ik} \ . 
\label{29}
\end{equation} 
The quantity $\frac{u ^{a}}{T}$ is a CKV of the optical metric 
$\bar{g} _{ik}$.   
This clarifies in which sense the modification of the conformal symmetry is again a symmetry. 
The transition from a de Sitter phase to a FLRW period may be regarded as a specific non-equilibrium configuration which microscopically is characterized by an equilibrium distribution function and macroscopically by the conformal symmetry of an optical metric with a time dependent refraction index. 
In the following sections we demonstate, how 
the presented scenario may be related to other approaches. 
Firstly, we point out that it is equivalent to a phenomenological vacuum decay model and secondly, that it may be translated into an equivalent dynamics of a scalar field with a specific potential term. 

\section{Phenomenological vacuum decay}

Combining the Friedmann equation in (\ref{21}) with the Hubble parameter in (\ref{22}), we obtain the energy density $\rho $, which is constant   
for $a \ll a _{e} $. For 
$a \gg a _{e}$ the familiar dependence  
$\rho \propto a ^{-4}$ is recovered. 
The  temperature behaves as $T \propto \rho ^{1/4}$.  
The evolution of the universe starts in a quasistationary state with finite initial values of temperature and energy density at $a \ll a _{e}$ and approaches the standard 
radiation dominated universe for 
$a \gg a _{e}$. 
This scenario may alternatively be interpreted in the context of a decaying vacuum (cf. \cite{LiMa,GunzMaNe}). 
The energy density may be split into 
$\rho = \rho _{_{\left(v \right)}} + \rho _{_{\left(r \right)}}$, where  
\begin{equation}
\rho _{_{\left(v \right)}} = \frac{3 H _{e}^{2}}{2 \pi }m _{P}^{2}
\left[\frac{a _{e}^{2}}{a ^{2} + a _{e}^{2}} \right]^{3}\ ,\quad 
\rho _{_{\left(r \right)}} = 
\left(\frac{a}{a _{e}} \right)^{2}\rho _{_{\left(v \right)}}\ .
\label{31}
\end{equation}
We have replaced here $G$ by the Planck mass $m _{P}$  according to  
$G = 1 /m _{P}^{2}$. 
The part $\rho _{_{\left(v \right)}}$ is finite for $a \rightarrow 0$ and decays as $a ^{-6}$ for $a \gg a _{e}$, while 
the part $\rho _{_{\left(r \right)}}$ describes relativistic matter with 
$\rho _{_{\left(r \right)}}\rightarrow 0$ for $a \rightarrow 0$ and 
$\rho _{_{\left(r \right)}}\propto a ^{-4}$  for $a \gg a _{e}$. 
The balance equations for $\rho _{_{\left(r \right)}}$ and  
$\rho _{_{\left(v \right)}}$ are 
\begin{equation}
\dot{\rho }_{_{\left(r \right)}} + 4H \rho _{_{\left(r \right)}} 
= - \dot{\rho }_{_{\left(v \right)}}\ ,\quad 
\dot{\rho} _{_{\left(v \right)}} 
+ 3H \left(\rho _{_{\left(v \right)}} 
+ P _{_{\left(v \right)}}\right) = 0 \ ,
\label{33}
\end{equation}
where  $P _{_{\left(v \right)}}/ \rho _{_{\left(v \right)}} 
= \left(a ^{2} - a _{e}^{2} \right)/ 
\left(a ^{2} + a _{e}^{2} \right)$. 
For $a \ll a _{e}$ the effective pressure $P _{_{\left(v \right)}}$ approaches 
$P _{_{\left(v \right)}} = - \rho _{_{\left(v \right)}}$. 
Effectively, this component behaves as a vacuum contribution.  
Acording to (\ref{33}) the radiation component may be regarded as emerging from the decay of the initial vacuum.  

\section{Scalar field dynamics}
 
In this section we discuss the interconnection between the previous fluid picture and the dynamics of a scalar field in terms of which  inflationary cosmology is usually described. 
To this purpose we start with the familiar identifications 
$\rho + P = \dot{\phi }^{2}$ and    
$\rho - P = 2 V \left(\phi  \right)$.
Integration yields  
\begin{equation}
\phi - \phi _{0} =  \frac{m _{P}}{\sqrt{2 \pi }}
{\rm Arsh} \frac{a}{a _{e}}
\quad\Rightarrow\quad 
\frac{a}{a _{e}} = \sinh \left[\sqrt{2 \pi }
\frac{\phi - \phi _{0}}{m _{P}}\right]\ ,
\label{39}
\end{equation}
where we have restricted ourselves to $\phi \geq \phi _{0}$.  
The potential is given by \cite{ZiBa3,MaaTayRou}
\begin{equation}
V = \frac{H _{e}^{2}}{2 \pi }m _{P}^{2}
\left[\frac{a ^{2}}{a _{e}^{2}} + 3 \right]
\left[\frac{a _{e}^{2}}{a ^{2} + a _{e}^{2}} \right]^{3} 
= \frac{H _{e}^{2}}{2 \pi }m _{P}^{2}
\frac{3 - 2 \tanh ^{2}\left[ \sqrt{2 \pi }
\frac{\phi - \phi _{0}}{m _{p}}\right]}
{\cosh ^{4}\left[\sqrt{2 \pi }
\frac{\phi - \phi _{0}}{m _{p}}\right]} 
\ . 
\label{40}
\end{equation}
A scalar field description with this potential implies the same cosmological dynamics as a self-interacting gas model in which the particles 
self-consistently move under the influence of an effective one-particle force 
characterized by the second equation in (\ref{23}). 
The above relations allow us to change from the fluid to the scalar field picture and vice versa. 
In principle, it is possible to ``calculate'' the potential if the self-interacting fluid dynamics is known. 

\section{Conclusions}

We have presented an exactly solvable gas dynamical model of a  
deflationary transition from an initial de Sitter phase to a subsequent radiation dominated FLRW period. 
The entire transition dynamics represents a specific non-equilibrium configuration of a self-interacting gas. 
Although connected with entropy production, this configuration is characterized  by an equilibrium distribution function of the gas particles.  
Mapping the non-equilibrium contributions onto an effective refraction 
index $n _{r}$ of the cosmic matter,  
the deflationary transition appears as  
the manifestation of a timelike conformal symmetry of an optical metric 
$\bar{g}_{ik}=g _{ik} + \left[1-n _{r}^{-2} \right]u _{i}u _{k}$ 
in which the refraction index $n _{r}$ changes smoothly from a very large value in the de Sitter period to unity in the FRLW phase. 
The deflationary gas dynamics may alternatively be interpreted as a production process of relativistic particles out of a decaying vacuum. 
Furthermore, there exists an equivalent scalar field description with an exponential type potential which generates the same deflationary scenario as the self-interacting gas model. \\
\\
{\bf Acknowledgment}\\
This work was supported by the Deutsche Forschungsgemeinschaft.

\end{document}